# AdaMR: Adaptable Molecular Representation for Unified Pre-training Strategy

**Yan Ding[1]**

Jiangsu Institute of Sports and Health

**Hao Cheng[1]**

Jiangsu Institute of Sports and Health

**Ziliang Ye[2]**

Southeast University

**Ruyi Feng[3]**

Southeast University

**Wei Tian[2]**

Southeast University

**Peng Xie[2]**

Southeast University

**Juan Zhang[1,4]**

Southeast University

**Zhongze Gu[1,2]\***

Southeast University

# Abstract

We propose **Ad**justa**b**le **M**olecular **R**epresentation (AdaMR), a new large-scale uniform pre-training strategy for small-molecule drugs, as a novel unified pre-training strategy. AdaMR utilizes a granularity-adjustable molecular encoding strategy, which is accomplished through a pre-training job termed molecular canonicalization, setting it apart from recent large-scale molecular models. This adaptability in granularity enriches the model's learning capability at multiple levels and improves its performance in multi-task scenarios. Specifically, the substructure-level molecular representation preserves information about specific atom groups or arrangements, influencing chemical properties

[1] Jiangsu Institute of Sports and Health, Nanjing 211112, China.

[2] State Key Laboratory of Digital Medical Engineering, School of Biological Science and Medical Engineering, Southeast University, Nanjing 210096, China

[3] School of Transportation, Southeast University, Nanjing, 210096, China

[4] Key Laboratory of Environmental Medicine Engineering, Ministry of Education, School of Public Health, Southeast University, Nanjing, 210009, China

Email: gu@seu.edu.cn

and functionalities. This proves advantageous for tasks such as property prediction. Simultaneously, the atomic-level representation, combined with generative molecular canonicalization pre-training tasks, enhances validity, novelty, and uniqueness in generative tasks. All of these features work together to give AdaMR outstanding performance on a range of downstream tasks. We fine-tuned our proposed pre-trained model on six molecular property prediction tasks (MoleculeNet datasets) and two generative tasks (ZINC250K datasets), achieving state-of-the-art (SOTA) results on five out of eight tasks.

# Introduction

The process of developing a new drug can take several decades, from its initial discovery to its commercialization. It is an extremely complex and drawn-out procedure. However, recent advancements in small-molecule computer modeling and analysis have revolutionized this process, dramatically reducing drug discovery timelines [1, 2]. This achievement is primarily due to the capability of these computer models to efficiently explore and generate novel molecular compositions based on known molecules [3, 4]. These models allow researchers to identify rational molecular structures previously absent from their molecular libraries.

A critical step in the functionality of computer models involves learning the representation of small molecules, making them identifiable and computable by computers. Before utlization by computer models, the string sequence encodings of molecules must be converted into numerical symbols. A molecular fingerprint encodes small molecules into binary vectors based on pre-defined rules [5]. Despite their ability to represent the presence of substructures in the molecules, fingerprints suffer from bit collision and vector sparsity, limiting their representation power. Subsequently, researchers tend to use deep learning models to obtain continuous vector representations of molecule sequences in a low-dimensional dense space to address issues like bit collision and vector sparsity [6-10]. Recent advancements in methodologies for fine-tuning large pre-trained models, utilizing architectures such as BERT [11], T5 [12], and GPT [13] for molecular modeling, have shown remarkable performance enhancements across various domains, including drug discovery. These methods obtain the embedded representation of molecules and accomplish multiple downstream tasks, as exemplified by works like PanGu [14].

Simplified Molecular Input Line Entry System (SMILES) [15] is often utilized as an original molecule sequence representation. This method employs a series of characters to denote the atoms and chemical bonds within a molecule. SMILES can depict highly complex chemical structures with a minimal number of characters, making it exceedingly valuable in cheminformatics and drug design. Additionally, SMILES supports the representation of stereochemistry, such as distinguishing chiral centers. A particular molecule can often be represented by multiple synonymous SMILES notations, which we refer to as generic SMILES; for example, CCO, OCC, and C(O)C all denote the structure of ethanol. Existing research on learning SMILES embeddings largely overlooks this synonymy. Works such as ChemBERTa [16], SmilesTransformer [17], and Knowledge-based BERT [18] typically train on single SMILES representations of molecules or employ canonical SMILES generated through specific algorithms to obtain unique identifiers for each molecule. However, disregarding SMILES synonymy results in distinct embeddings for the same molecule, and reliance on single SMILES representation fails to comprehensively capture the relationships and conversions between SMILES's syntax and molecular structure, leading to substantial loss of semantic information and an inability to construct a high-quality chemical space.

Sequence encoding methods also exhibit variations in granularity. Some methods directly split molecules into atomic granularity [19-21], while others opt for functional groups as basic tokens [22]. In the field of biochemistry, the partitioning of molecular linear representations at different granularities embodies various levels of structural information within molecules; hence, sequence encoding should not be simplistically viewed as a mapping from characters to numbers. Instead, it necessitates a nuanced understanding and handling of the hierarchical structural details inherent in these molecular representations. Existing pre-trained large models traditionally utilize a fixed granularity for encoding small molecules, overlooking the diversity of molecular structure representations. For instance, XMOL [19] designs a generative molecular pre-training model using an atomic-level encoder. While it performs well in generative tasks due to its task design, its performance is found to be mediocre in property prediction tasks, and it does not consistently achieve state-of-the-art (SOTA) results across a variety of downstream tasks. Therefore, a single molecular encoding method and a universal molecular structure representation may not be widely applicable to different downstream tasks and may cause information loss during the encoding process. These limitations make it unsuitable for practical multi-target production environments. Additionally, Group SELFIES [22], utilizing substructure-level representation encoding, enhances the model's ability to learn representation distributions on generic molecular datasets. However, currently, there is a lack of extensive research investigating the application of pre-trained models based on substructure-level encoding to downstream tasks in the field.

To establish a uniform pre-training model that can address the encoding defects in computer modeling for small molecules, we need to make sure that there is as little information loss as possible throughout the encoding process in order to achieve optimal outcomes for a wide range of downstream tasks. Our study proposes a granularity-adjustable encoding method for small molecules that learns the distributions of molecular representations at various granularities in a unified manner during the pre-training phase. We apply different encoding granularities to different downstream tasks. Different levels of encoding granularity encapsulate distinct tiers of chemical structural information within drug molecules. Simultaneously, downstream tasks exhibit varying preferences for such information, essentially reflecting the tasks' reliance on different encoding granularities. Additionally, we design a pre-training task named molecular canonicalization. This task, which generates canonical SMILES from generic SMILES, enables the model to deeply understand the intrinsic correlation features within molecular sequences while enhancing the model's capability to perform generative tasks. The unique design of this task also enables the model to learn the rich intrinsic positional correlation information within SMILES representations. We validate its effectiveness through downstream tasks such as classification, regression, and generation.

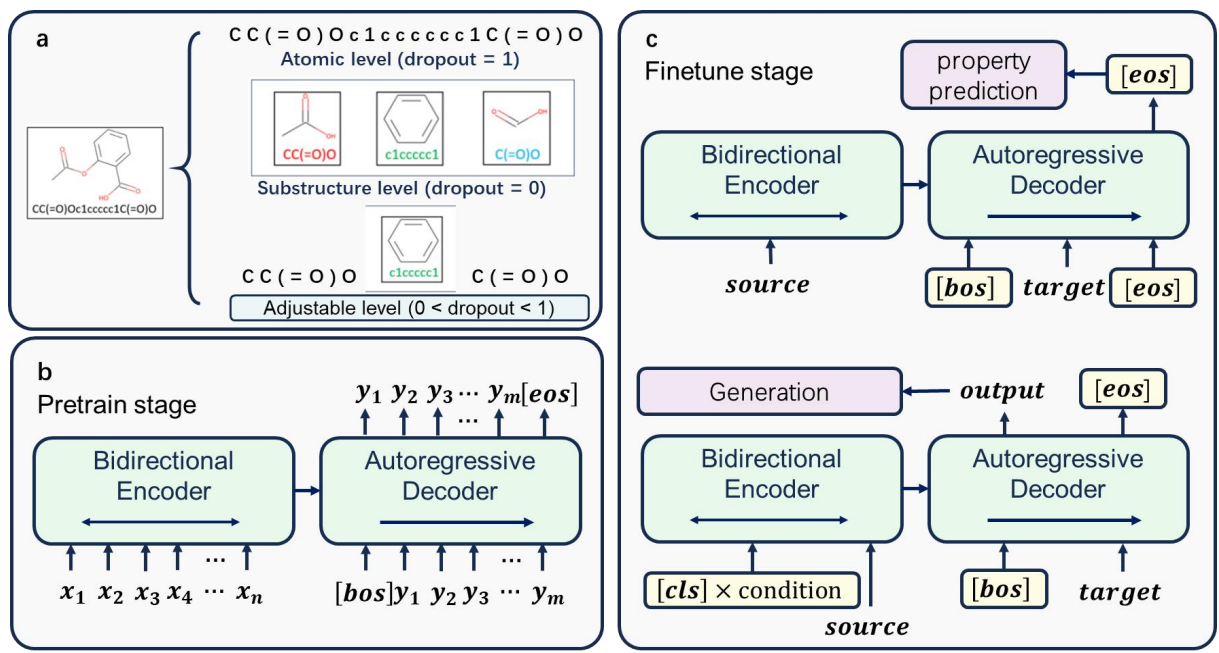

**Figure 1**: The process and structure of AdaMR. a) Encoding a molecule with different granularities. Atomic-level encoding involves breaking down the molecule into individual atoms. Substructure-level encoding means breaking down the molecule into substructures that retain fundamental characteristics, such as a benzene ring. Our encoding method utilizes dropout to mix atomic-level encoding and substructure-level encoding according to a certain probability during the pre-training phase. b) Pre-training an encoder-decoder Transformer model on molecular canonicalization tasks. The input is an encoded generic SMILES, and the output is an encoded canonical SMILES. c) Fine-tuning the pre-trained model on a range of downstream tasks. For downstream tasks including property prediction and molecule generation, we devise different fine-tuning strategies based on the pre-trained model to ensure the comprehensive utilization of AdaMR's advantages.

## Materials and Methods

In general, we present a novel molecular pre-training method that incorporates adjustable molecular encoding and a dedicated molecular canonicalization task to train a Transformer model with an encoder-decoder architecture. **Figure 1** illustrates the overall structure of AdaMR. Model pre-training is conducted using the comprehensive ZINC20 dataset [23]. The adjustable molecular encoding comprises two primary modes: atomic-level encoding and substructure-level encoding. During the pre-training phase, SMILES representations of molecules are encoded using substructure-level encoding. There is a 20% chance that the encoded substructure-level tokens will degrade to atomic-level tokens. This approach enables the pre-training model to learn both atomic and substructural information. During the fine-tuning phase, for property prediction and molecule generation tasks, the pre-trained model is fine-tuned and evaluated using substructure-level encoding. For molecule generation tasks, the pre-trained model is fine-tuned and evaluated using atomic-level encoding.

## Granularity-adjustable Molecular Tokenizer

To maximize the utilization of feature information at both the atomic and substructural levels within the pre-training model, we introduce a novel approach called "granularity-adjustable molecular tokenization." This method initiates by establishing a vocabulary of high-frequency SMILES substrings through extensive mining of a large chemical dataset. The process iteratively counts the occurrences of all token pairs in the tokenized atomic-level SMILES, merging the most frequently occurring token pair into a new token and incorporating it into the vocabulary. This iterative step continues until one of the following conditions is met: (1) the desired vocabulary size is achieved, or (2) no more token pairs exceed a specified frequency threshold. The maximum vocabulary size and frequency threshold serve as hyperparameters in the training of the vocabulary. Subsequently, the trained vocabulary is utilized for the tokenization of SMILES in deep learning models. During tokenization, each SMILES string undergoes initial substructure-level tokenization. Then substructure-level tokens are decomposed into atomic-level tokens following a specific probability, which is called spe dropout. The capacity of the proposed tokenization method to capture information at a granularity beyond the atomic level is analogous to the relationship between phrases and words in natural language.

## Transformer Backbone

A popular artificial intelligence model based on deep learning, the Transformer model [24] is used to handle sequential data in a variety of fields, including natural language processing, computer vision, and biomolecular sequence analysis. Transformers can process long sequential data and are versatile in tasks such as generation and classification. The whole structure of the Transformer model is an encoder and a decoder. The encoder encodes the input sequential data to obtain a high-dimensional vector representation, and the decoder decodes this high-dimensional vector to generate an output. Given the diversity of our model's downstream tasks, involving both molecule generation and property prediction, an encoder-decoder architecture is adopted. This architecture excels at comprehending complex inputs and providing relevant outputs [25]. The tasks addressed by the model will be discussed in the following two sections.

The core of our model is a Transformer encoder-decoder structure, comprising 12 layers, with each layer incorporating 12 attention heads. The hidden size is set to 768, and the dropout rate is set to 0.1. We employ learnable absolute position embedding. This configuration enhances the extraction of key features from molecular SMILES. To cater to various downstream tasks, special modules are added after the output of the decoder. These modules process the output of the decoder and convert it into a label suitable for each downstream task.

## Pre-training Strategy

We anticipate that the pre-training task will enable the model to thoroughly grasp the syntax of SMILES and its corresponding relationship with molecular structure. Consequently, it is essential to take into account the identical semantic information contained within multiple generic SMILES representations of a specific molecule and to effectively learn the knowledge encoded at varying levels of granularity. Therefore, we design a pre-training task named molecular canonicalization to unify these objectives. Specifically, for one molecule, we first use RDkit [26] to obtain multiple generic SMILES representations and a unique canonical SMILES representation of the molecule. Then, we select N generic SMILES and pair them with the canonical SMILES to form N sample pairs, which serve as the training data. The training objective of the model is to input generic SMILES and output canonical SMILES. The pre-training strategy not only meets the aforementioned objectives but also expands the data by introducing different equivalent SMILES representations for molecules, allowing the model to fully learn the intrinsic information of the SMILES. Due to resource constraints, we heuristically generate three generic SMILES strings for pairing with each canonical SMILES string.

We conduct a series of statistical analyses and processing steps on the ZINC20 dataset to determine the pre-training hyperparameters. To ascertain the distribution of sequence lengths, we first examine the length of each of the 1.87 billion molecular data points .We then adjust the maximum sequence length to ensure that it encompasses 99% of the molecules in the dataset. 1% of excessively long molecular sequences are gathered as a separate dataset for future independent training, which saves GPU memory and reduces computational complexity.

Additionally, during the pre-training phase, we introduce a substructure dropout mechanism within the tokenizer. During tokenization, there is a certain probability of inserting atomic-level encoding tokens in the encoding tokens at the substructure-level. The dynamic dropout strategy is pivotal in enabling the pre-training process to simultaneously acquire information at both atomic and substructure levels. When deploying the model for downstream tasks, we adjust the dropout probability, setting it to either 0 or 1, depending on the specific requirements of the task at hand. The adaptive approach ensures optimal model performance across a range of downstream tasks.

## Downstream Tasks

To verify the effectiveness of the proposed molecular encoding strategy and pre-training task and to validate the performance of the proposed pre-training model, an extensive series of experiments are conducted across a range of downstream tasks. These tasks include molecular property prediction and molecule generation. With the aim of obtaining canonicalized SMILES representations and reducing the complexity of sampling space in downstream molecule generation tasks, pre-training of the model is conducted using an N-to-1 SMILES-alias conversion task. Subsequential fine-tunings are conducted to complete the aforementioned downstream tasks.

## Molecular Property Prediction

In terms of property prediction, we differentiate between molecule regression tasks (predicting a continuous output variable) and molecule classification tasks (predicting a categorical output variable). Despite sharing the same initial processing steps, these downstream tasks primarily differ in the dimensions of the output layers. The SMILES tokens are input into the encoder, with the start position of SMILES tokens marked with "[bos]" and the end position with "[eos]" before being fed into the decoder. Subsequently, a feature vector is extracted at the last position of the autoregressive process. This feature vector is then fed into the downstream task head for property prediction. Both regression tasks and classification tasks involve task heads comprising two fully connected layers. For the output dimension of the task head, it is set to 1 for regression tasks and to the number of classes for classification tasks. For classification tasks, we adhere to the MoleculeNet benchmarks by employing cross-entropy loss during training and subsequently assessing performance using the ROC-AUC score. For regression tasks, we employ mean squared error (MSE) loss during training and evaluate model performance using root mean squared error (RMSE).

## Molecule Generation

Following the design of the experiments and dataset creation method of XMOL [19], we conduct two generative tasks based on ZINC250k, which are Distribution learning-based Generation (DG) and Goal-directed molecule Generation (GG).

**Distribution Learning-based Generation (DG)**

The DG task aims to train the model to learn the distribution of molecular representations in the dataset and generate molecules that conform to this distribution. During the training phase, the encoder takes a [cls] token, and the decoder takes the starting position of SMILES with a [bos] token, predicting the molecular representation in an autoregressive manner. Random sampling is employed during the generation phase. Validity, uniqueness, and novelty are used as evaluation metrics. In molecule generation, validity assesses whether the generated molecules adhere to predefined chemical rules and properties. Valid molecules should conform to standard chemical bonding and follow principles of substance stability and feasibility. A model demonstrating validity can generate molecular structures that comply with real-world chemical constraints. Uniqueness evaluates whether the generated molecular structures are significantly different from known molecules. In molecule generation tasks, pursuing uniqueness means generating new molecules rather than simply replicating or modifying known molecular structures, which contributes to enhancing the innovativeness of the generated structures. Novelty emphasizes whether the generated molecules are entirely new compared with known databases or literature, as opposed to molecules that have already been extensively studied or discovered. Novelty is a crucial objective in molecule generation tasks because it drives models to produce molecules with advancement and originality, expanding the current molecule database.

**Goal-directed Molecule Generation (GG)**

The GG task is designed to provide conditions to guide the molecule generation task. The only thing that seperates this task's procedure from the DG task is the addition of condition information to the [cls] token input for the encoder. The [cls] embedding vector is modified by multiplying with the value of the generation goal to obtain a new [cls] vector. For this task, we use random sampling for molecule generation. The Quantitative Estimate of Drug-likeness (QED) value is specifically chosen as the optimization target. The evaluation process includes determining the number of molecules that satisfy the condition within the top 5 molecules after sorting the generated molecules according to the specified condition.

# Ablation Experiment

We design two sets of ablation experiments: an ablation study on the pre-training tasks and another one examining the impact of different encoding granularities on the model's performance in downstream tasks. This allows us to further validate the efficacy of our proposed molecular encoding and the molecular canonicalization pre-training task in constructing a more effective chemical space for downstream tasks, and to investigate the downstream tasks' reliance on different molecular encoding granularities.

**Ablation Experiments on the Pre-training Task**

This part of the experiments aims to confirm that the proposed molecular canonicalization pre-training task improves performance on downstream tasks. In this setup, we employ a novel 1-to-N pre-training task where canonical SMILES serve as inputs, with the expected output being diverse generic SMILES representations representing the same molecular structure, thereby leveraging the synonymity of generic SMILES for one specific molecule. For the model pre-trained on this 1-to-N task, we employ atom-level molecular encoding, then fine-tune the model on all downstream tasks using the previously mentioned fine-tuning methods. The performance of the model is assessed using the same evaluation methods applied to the respective downstream tasks.

**Ablation Experiments on Encoding Granularity**

This part of the experiments is designed to examine how different levels of encoding granularity influence the performance during fine-tuning on downstream tasks and to explore the preferences of various downstream tasks regarding the encoding granularity. The model pre-trained for the molecular canonicalization task is fine-tuned according to the aforementioned method for the downstream tasks. Subsequently, we evaluate the performance of the model using the consistent assessment methodologies applied to each of these tasks.

# Data Collection

ZINC20 is an expansive, publicly accessible database housing an astonishing assembly of around 1.87 billion meticulously annotated, commercially viable molecules. It emerges as a monumental and invaluable resource, purposefully designed to catalyze advancements in the realm of ligand discovery. For pre-training strategy, the compact and information-rich SMILES encodes the structural and chemical properties of molecules in a human-readable format. In this work, we use SMILES representations of all molecules in ZINC20 for the pre-training phase.

For downstream tasks, molecular property prediction holds significant importance in the field of drug discovery. MoleculeNet serves as a widely recognized benchmark for evaluating the performance of molecular property prediction models. It encompasses a collection of datasets that capture various molecular properties. To enable a comprehensive comparison with the baseline models, we followed X-MOL [19] and selected six commonly used sub-datasets from MoleculeNet. Out of these, three datasets were utilized for molecular property classification tasks, namely BACE, BBBP, and ClinTox. Additionally, three datasets were chosen for molecular property regression tasks, namely ESOL, FreeSolv, and Lipophilicity. The details of these datasets are as follows:

**BACE:** The BACE dataset contains 1522 molecules, each accompanied by binary labels for classification tasks that indicate their binding affinity with human β-secretase 1 (BACE-1). These labels are derived from experimental values reported in scientific literature over the past decades and are utilized for classification tasks.

**BBBP:** The BBBP dataset contains over 2000 compounds, each accompanied by binary labels for classification tasks that indicate their permeability characteristics across the blood-brain barrier. The data originates from a research study focused on modeling and predicting barrier permeability.

**ClinTox:** The ClinTox dataset contains two distinct sub-datasets designed for classification tasks. It serves as a comparative analysis between FDA-approved drugs and drugs that encountered clinical trial failures due to toxicity. The dataset comprises 1491 known drug structures, each associated with binary labels indicating drug toxicity and FDA approval status.

**ESOL:** The ESOL dataset comprises a collection of 1128 compounds and serves as the basis for a regression task aimed at estimating the solubility in water using the compounds' SMILES representation.

**FreeSolv:** The FreeSolv dataset contains over 600 compounds, each accompanied by a hydration-free energy value in water. These values serve as the labels for a regression task aimed at predicting solvation energies.

**Lipophilicity:** The Lipophilicity dataset contains 4200 compounds derived from the ChEMBL dataset, each accompanied by experimental data on octanol/water partition coefficients. These data points are employed for regression tasks.

The scaffold split processing method from MoleculeNet is followed for all six datasets. Each dataset is divided into training and test sets in an 8:2 ratio. Data augmentation is performed before training because of the relatively small size of these datasets. The augmentation method involves generating different valid SMILES representations for each molecule in the training set using the RDkit tool and incorporating these representations into the training set. Further details and usage of the datasets are provided in **Table 1.**

**Table 1**: The scale and role of the dataset used in the experiment

| Dataset  | Samples amount | Usage          |
|----------|----------------|----------------|
| ZINC20   | 1.87billion    | Pre-training   |
| ZINC250K | 249455         | Generation     |
| BACE     | 1513           | Classification |
| BBBP     | 2039           | Classification |
| Clintox  | 1478           | Classification |
| ESOL     | 1128           | Regression     |
| **FreeSolv** | 642 | Regression |
| **Lipophilicity** | 4200 | Regression |

The ZINC250K dataset is employed for molecule generation tasks. The ZINC250K database is a curated collection designed to represent commercially available chemical compounds for virtual screening. ZINC250K contains around 250,000 compounds selected based on their "lead-likeness" or "drug-likeness," indicating properties that render molecules suitable for further development in the drug discovery process.

## Experimental Environment

PyTorch framework with Cuda 11.6 is utilized to implement the entire pre-training model and specific downstream tasks. Each downstream task with pre-training is conducted on one Tesla V100 GPU (32GB) for simultaneous model training. The pre-training stage utilizes 16 GPUs and requires around 240 hours.

# Result

## Classification of Molecular Property

In property prediction, tasks generally fall into two categories: classification and regression. Models tailored for these tasks produce outputs with different dimensions in their task headers, and they employ distinct evaluation metrics.

For the classification tasks, we assess model performance using three subtasks from the MoleculeNet dataset: BACE, BBBP, and ClinTox. Since our overall pre-training architecture is an encoder-decoder, in downstream property prediction tasks, we mimic the training method of the pre-training task. The encoder inputs SMILES, and the decoder inputs SMILES with [bos] at the beginning and [eos] at the end. We extract the decoder output vector corresponding to the position of [eos] as the input for the classification head. For the classification head, following Huggingface's implementation approach for downstream task heads of pre-trained models, we add two fully connected layers with a tanh activation function in between. We utilize substructure-level encoding in the task, and the evaluation metric employed is the ROC-AUC score, where a higher score indicates better performance. **Figure 2** illustrates the results and comparisons with existing research. We use nine models in MolecularNet as baselines to evaluate our model's performance while also comparing it with the best-performing Pangu model.

In molecular property classification tasks, the performance of the nine baseline models on the three datasets does not achieve higher ROC-AUC scores compared to our model and the existing state-of-the-art (SOTA) model. Our model achieves ROC-AUC scores of 0.917 and 0.894 on the BBBP and BACE datasets, respectively, approaching the current SOTA results achieved by Pangu. On the ClinTox dataset, the ROC-AUC score is 0.969, surpassing PanGu's 0.953. In the experiments on the three datasets, the performance and stability of our proposed model are significantly improved compared to the nine baseline models.

## Regression of Molecular Property

For the regression tasks, we evaluate model performance using three subtasks from the MoleculeNet dataset: ESOL, FreeSolv, and Lipophilicity. The structure of the downstream task head remains consistent with that used for classification tasks with continuous labels. The encoding method for SMILES is set to substructure-level. The evaluation metric employed is RMSE, where a lower score indicates better results. **Figure 2** illustrates the results and provides comparisons with existing research. We use eight models in MolecularNet as baselines while comparing our model's performance with that of Pangu.

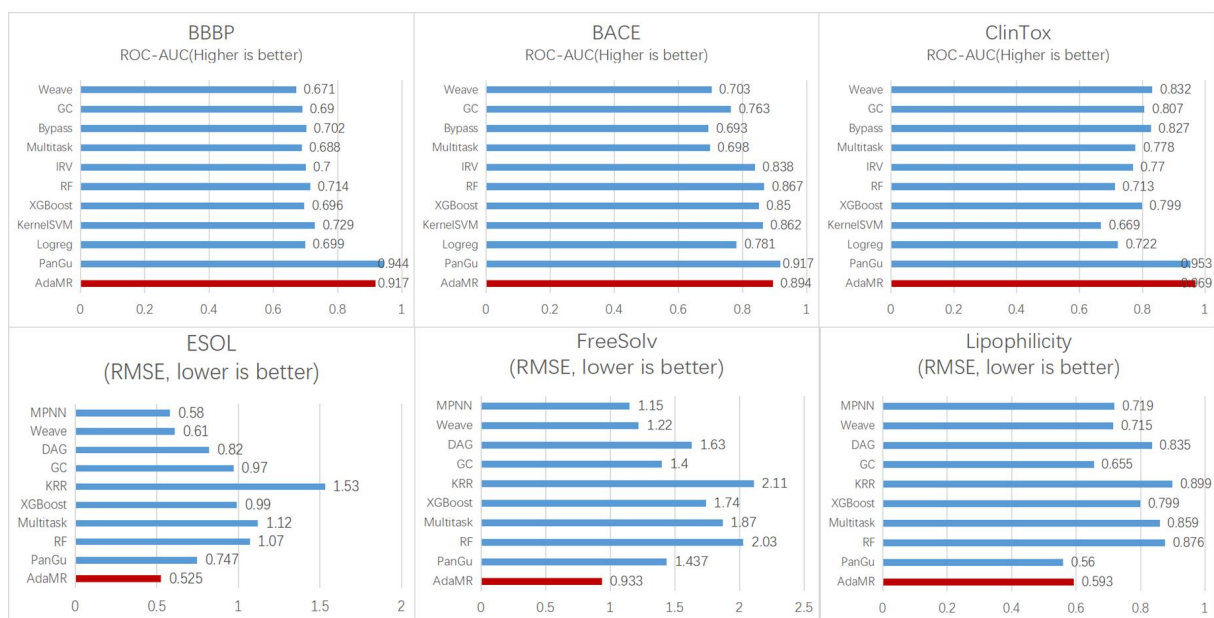

**Figure 2:** Comparisons of the performance of our proposed model and a range of benchmark models in downstream tasks of molecular property prediction.

In regression tasks, AdaMR achieves RMSE scores of 0.525, 0.933, and 0.593 on the ESOL, FreeSolv, and Lipophilicity datasets, respectively. AdaMR outperforms the SOTA results to varying degrees on each dataset. Notably, on both the ESOL and FreeSolv datasets, the Pangu model performs even worse than the best-performing baseline model on both datasets.

Our experimental results demonstrate the superiority of AdaMR. Among all six property prediction tasks, three achieve SOTA results. On the ClinTox dataset, there is a 1.6% increase in ROC-AUC compared to the best existing result. On the ESOL dataset, the RMSE decreases by 9.48% relative to the best result. On the FreeSolv dataset, we observe an 18.87% reduction in RMSE. In the other three tasks, we achieve metrics that are closest to the best results. It is noteworthy that while PanGu achieves SOTA results in classification tasks, its performance in regression tasks is suboptimal. In contrast, AdaMR demonstrates strong performance, achieving results close to or surpassing SOTA in both classification and regression tasks. This highlights the robust performance and adaptability of AdaMR across a range of downstream tasks.

# Distribution Learning-based Generation (DG)

Molecule generation tasks can be divided into conditional and unconditional generation tasks. Unconditional generation can be seen as sampling from the output distribution of a pre-trained model to obtain molecules. Molecules generated in this way are consistent with the distribution of the training data, and thus this method can be referred to as distribution learning-based generation.

For DG, we use the original ZINC250K dataset and randomly split it into training and test sets, with a ratio of approximately 8:2. In terms of the model, we employ a standard pre-trained model without adding any additional modules. The encoder inputs SMILES with [cls] added at the beginning, while the decoder inputs SMILES with [bos] at the beginning and [eos] at the end. The encoding method for SMILES is set to atomic-level. Random sampling is used to sample tokens of molecules from the distribution output by the decoder. The experimental results and comparison with existing research are shown in **Table 2**.

**Table 2**: Comparison of results for DG task

| Method   | Validity | Uniqueness | Novelty |
|----------|----------|------------|---------|
| JT-VAE   | -        | 100%       | 100%    |
| GCPN     | 20%      | 99.97%     | 100%    |
| MRNN     | 65%      | 99.89%     | 100%    |
| GraphNVP | -        | 94.89%     | 100%    |
| GraphAF  | 68%      | 99.10%     | 100%    |
| X-MOL | 85.28% | 99.91% | 100% |
| AdaMR | 90.7% | 99.1% | 93.2% |

In the generation task, DG aims to learn molecular distributions from training data and evaluate model performance by randomly sampling molecules from the model output distribution during the inference phase. We assess model performance using validity, uniqueness, and novelty. These characteristics are interrelated in molecular design and evaluation. Validity ensures the practicality and efficiency of a molecule for its intended use, while uniqueness provides new properties distinct from existing molecules or drugs, and novelty drives scientific and technological advancement. Particularly in the competitive field of drug development, these traits are crucial for ensuring a molecule's market success. AdaMR enhances the validity of generated molecules while ensuring that uniqueness and novelty are above 90%. Such assessment outcomes demonstrate the high reliability of molecules generated by AdaMR. Overall, validity, novelty, and uniqueness are indeed interconnected. Low validity suggests that the model has not fully understood the sequence structure information contained in the training data. On the other hand, molecules generated by the model tend to have more chaotic structures, which can indeed increase novelty and uniqueness metrics, but this is not our ultimate goal.

## Goal-directed Molecule Generation

Conditional generation tasks involve guiding the model to generate molecules that meet specific criteria. In our study, we conduct the GG task based on the ZINC250K dataset, with a focus on optimizing the Quantitative Estimation of Drug-likness (QED) value.

QED serves as the optimization goal, representing a quantitative measure of a compound's drug-likeness. This metric considers various chemical properties such as molecular weight, lipophilicity, polarity, saturation, and aromaticity, aiming to assess a molecule's suitability as a drug candidate. QED scores typically range between 0 and 1, with higher scores indicating greater drug-likness.

To adapt the pre-trained model for this task, we make some slight adjustments. We multiply the embedding feature vector corresponding to the [cls] token by the QED attribute value of SMILES and input it into the encoder. The decoder inputs SMILES with [bos] and [eos] added at the beginning and end, respectively. Since the maximum QED value of the molecules in the ZINC250K dataset is 0.948, we use this value for molecule generation during the evaluation phase. The SMILES encoding method is set to atomic-level encoding. The evaluation results are presented in **Table 3**.

**Table 3**: Comparison of the top 3 QED values for molecules generated in the GG task

| Method | 1st | 2nd | 3rd |
|---|---|---|---|
| ZINC(Dataset) | 0.948 | 0.948 | 0.948 |
| JT-VAE | 0.925 | 0.911 | 0.910 |
| GCPN | 0.948 | 0.947 | 0.946 |
| MRNN | 0.844 | 0.947 | 0.946 |
| GraphAF | 0.948 | 0.948 | 0.947 |
| X-MOL | 0.948 | 0.948 | 0.948 |
| AdaMR | 0.948 | 0.948 | 0.948 |

**Table 4**: The top 10 molecules with the highest QED values generated by AdaMR in the GG task, with QED = 0.948 as the goal.

| TOP 10 | QED |
| --- | --- |
| Cc1nn(C)c(Cl)c1COc1ccc2c(c1)CCC(=O)N2 | 0.9484 |
| Cc1csc([C@H](C)NC(=O)[C@@H]2COc3ccccc3C2)n1 | 0.9482 |
| O=C(NC[C@H]1COc2ccccc2O1)c1ccccc1Cl | 0.9482 |
| CCc1nn(C)cc1C(=O)Nc1nc(C(F)(F)F)cs1 | 0.9482 |
| Cc1nnc(CNC(=O)N2CCC[C@@H]2c2ccccc2)s1 | 0.9481 |
| O=C(NC[C@@H]1COc2ccccc2O1)[C@@H]1CCc2ccccc21 | 0.9480 |
| O=C(NCC1Oc2ccccc2O1)c1cc2c(s1)CCC2 | 0.9480 |
| Cc1cccc(C)c1NC(=O)CN1C(=O)COc2ccccc21 | 0.9480 |
| Cc1ccc(C)c(CC(=0)Nc2nc3c(s2)COCC3)c1 | 0.9479 |
| Cc1nc(NC(=O)N2CCC[C@H]2Cn2cccn2)sc1C | 0.9479 |

Table 4 shows the QED values of the top 10 generated molecules. The GG task aims to enable the model to generate molecules that meet specific criteria. After training, we use the optimal QED value of 0.948 found in the training data as a conditional input. The results indicate that molecules generated by AdaMR exhibit superior QED values. In the top 3 SMILES based on QED values, all molecules generated by AdaMR have QED values above 0.948, surpassing other models. Furthermore, 80% of the molecules in Table 4 have QED properties exceeding 0.948, a significant improvement compared to XMOL's 20%.

## Ablation on Different Encoding Granularities (Atomic-level and Substructure-level):

Table 5: Comparison of metrics for atomic-level encoding and substructure-level encoding in property prediction tasks based on different encoding granularities. For classification tasks, a higher ROC-AUC value indicates better performance. For regression tasks, a lower RMSE value indicates better performance.

| Task type | | AdaMR | |
| --- | --- | --- | --- |
| | | atomic-level | substructure-level |
| Classification (ROC-AUC) | BACE | 0.858 | 0.894 |
| | BBBP | 0.900 | 0.917 |
| | ClinTox | 0.961 | 0.969 |
| Regression | ESOL | 0.540 | 0.525 |

| (RMSE) | FreeSolv | 1.088 | 0.933 |
| | Lipophilicity | 0.776 | 0.593 |

Table 6: Comparison of metrics for atomic-level encoding and substructure-level encoding in DG tasks.

| Encoding Method | Validity | Uniqueness | Novelty |
| --- | --- | --- | --- |
| atomic level | 90.7% | 99.1% | 93.2% |
| substructure level | 18.3% | 100% | 99.8% |

Table 7: The top 10 molecules with the highest QED values generated by AdaMR encoding by substructure-level method, with QED=0.948 as the goal in the GG task.

| TOP 10 | QED |
| --- | --- |
| O=C1N[C@]2(CCCO2)CN1c1ccc(Cc2ccccc2)cn1 | 0.9482 |
| COc1ccc(Br)cc1NC(=O)c1ccncc1 | 0.9477 |
| CC(=O)C1CCN(C(=O)c2c[nH]c(-c3ccsc3)n2)CC1 | 0.9474 |
| C[C@@H](C(N)=O)N1CCN(c2ccccc2)c2ccc(Cl)cc21 | 0.9461 |
| Cc1ccc(N2CN[C@H](C3=Nc4ccccc4C3=O)C2)cc1 | 0.9459 |
| Cc1ccc(NC(=O)Cc2ccc(F)cc2Cl)c([O-])c1 | 0.9451 |
| C[C@@H](NC(=O)c1ccco1)C(=O)N1CCC[C@@H]1c1cccs1 | 0.9426 |
| Cc1c(Cl)cccc1S(=O)(=O)NCc1cccs1 | 0.9424 |
| [NH3+][C@H](c1cccc(Cl)c1Cl)N1CCC[C@H]1c1nncs1 | 0.9413 |
| C[NH+](Cc1ccc(Br)o1)[C@H]1CCCc2cccnc21 | 0.9413 |

We conduct ablation experiments on SMILES encoding methods to demonstrate the distinct roles of atomic-level and substructure-level information. Benefiting from the dynamic substructure dropout strategy employed during the pre-training phase, the pre-trained model sinultaneously acquires comprehension of both substructural and atomic levels of information. Under varying degrees of encoding granularity, the model yields reasonable outcomes across different downstream tasks.

Initially, we perform ablation experiments with molecular encoding granularities on the six previously mentioned molecular property prediction datasets, and the experimental results are shown in **Table 5**. For property prediction tasks, the results obtained with substructure-level encoding consistently outperform those with atomic-level encoding. Conversely, for generation tasks, the overall results of atomic-level encoding are superior to substructure-encoding. The properties of molecules are often determined by the charactersitcs of their functional groups and pharmacophores, with the molecule generation task requiring a focus on the atomic-level molecular skeletons. This explains why the different granularities of molecular encoding cause the different performance of the model on the two types of downstream tasks.

**Table 6** presents the results of the ablation experiments on the DG task for different encoding granularities**.** In downstream tasks of DG, substructure-level encoding achieves the highest level of novelty and uniqueness, but with low validity. This indicates that the model is currently not effective in handling the concatenation of substructures represented by SMILES, although it can explore multiple combinations of substructures. The practical significance of the generated molecules remains to be verified. The novelty and uniqueness obtained by atomic-level encoding are slightly lower than those obtained by substructure-level encoding, but the generated SMILES strings have 100% validity, indicating that the generation model can produce effective molecular skeletons by learning atomic-level molecular encoding. However, this also makes it easier for the model to learn the trends of molecular skeletons in the training data, resulting in the homogenization of generated molecules.

**Table 7** shows the top 10 molecules with the highest QED values generated by AdaMR encoding using substructure-level method with the goal of QED=0.948, representing the results of the ablation experiments on the GG task. For the GG task, substructure-level encoding yields the best batch of molecules with QED values deviating from the target values greater than atomic-level encoding. A larger encoding granularity results in the model's inability to learn atomic-level information that affects QED values, while a smaller encoding granularity helps the model achieve better performance in GG tasks.

## Ablation on Pre-training Tasks

To demonstrate the superiority of the pre-training task, we conduct an ablation experiment with a 1 to N pre-training strategy. We invert the molecular canonicalization task, pre-training the model to generate generic SMILES from canonical SMILES. Comparative experiments are conducted on generative tasks. **Tables 8 and 9** present the results of the ablation experiments on the DG and GG tasks.

**Table 8**: Ablation experiments on pre-training tasks. We train two models: one for the molecular canonicalization task, i.e., canonical SMILES to generic SMILES (N to 1), and another for the molecular generalization task, i.e., generic SMILES to canonical SMILES (1 to N). We set N to 3. We compare their performance on generation tasks. This table compares the performance of the DG task.

| Pre-training task | Validity | Uniqueness | Novelty |
| --- | --- | --- | --- |
| N to 1 | 90.7% | 99.1% | 93.2% |
| 1 to N | 84.7% | 99.8% | 95.7% |

In the 1 to N pre-training task, the model generates equivalent SMILES aliases based on canonicalized SMILES and samples the entire molecular space as much as possible by increasing the complexity of the sampling space. Therefore, when conducting downstream tasks of molecule generation, the high uniqueness and novelty of the molecules generated by the model pre-trained on 1 to N tasks are explainable. In the N to 1 pre-training task, the model generates canonicalized SMILES strings based on generic SMILES aliases, intending to reduce the complexity of the sampling space so that the generated strings are limited to the canonicalized SMILES space. Hence, a model pre-trained on N to 1 tasks can achieve higher validity scores in generation tasks. Due to the limited sampling space, the novelty and uniqueness scores obtained are not as good as the model pre-trained on 1 to N tasks. Consider the small gap between novelty and uniqueness scores brought by the two pre-training methods, and the markedly higher validity of SMILES generated by the model pre-trained on N to 1 tasks, we use the model pre-trained on N to 1 tasks for the downstream task of molecular generation.

**Table 9**: This table illustrates the performance of the 1 to N pre-training task in the GG task. It can be compared with Table 4.

| TOP 10 | QED |
| --- | --- |
| C[C@@H]1OCC[C@H]1C(=O)Nc1ccc(-n2ccc(Cl)n2)cc1 | 0.9483 |
| O=C(N[C@H]1CCCC[C@@H]1n1ccnc1)c1nccc1Cl | 0.9482 |
| COc1nccc1CNC(=O)N1Cc2ccccc2C[C@@H]1C | 0.9480 |
| COCc1cc(NC(=O)N2CCCc3sccc32)n(C)n1 | 0.9480 |
| COc1ccc(F)c(NC(=O)c2cc(Br)c[n-]2)c1 | 0.9479 |
| CCn1ccc(C(=O)N[C@@H]2CCSc3cccc(F)c32)n1 | 0.9479 |
| C[C@@H](c1ccco1)N(C)C(=O)N[C@@H]1CCCc2c1cnn2C | 0.9479 |
| Cc1ccc(NC(=O)[C@@H](C)N2CCc3sccc3C2)nc1 | 0.9479 |
| Cc1c(C(=O)Nc2ccc(OC(F)(F)F)cc2)cnn1C | 0.9478 |
| Cc1cc(C)n(-c2cccc(NC(=O)[C@@H]3C[C@H]4CC[C@@H]3O4)c2)n1 | 0.9477 |

**Figure 3**: Stuctures of molecules generated in different GG tasks. a) The top 3 molecular 3D structures with the highest QED values generated by AdaMR in the GG task, with QED = 0.948 as the goal. b) The top 3 molecular 3D structures with the highest QED values generated by AdaMR pre-trained on 1 to N tasks in the GG task, with QED = 0.948 as the goal. c) The top 3 molecular 3D structures with the highest QED values generated by AdaMR encoded by the substructure-level method in the GG task, with QED = 0.948 as the goal.

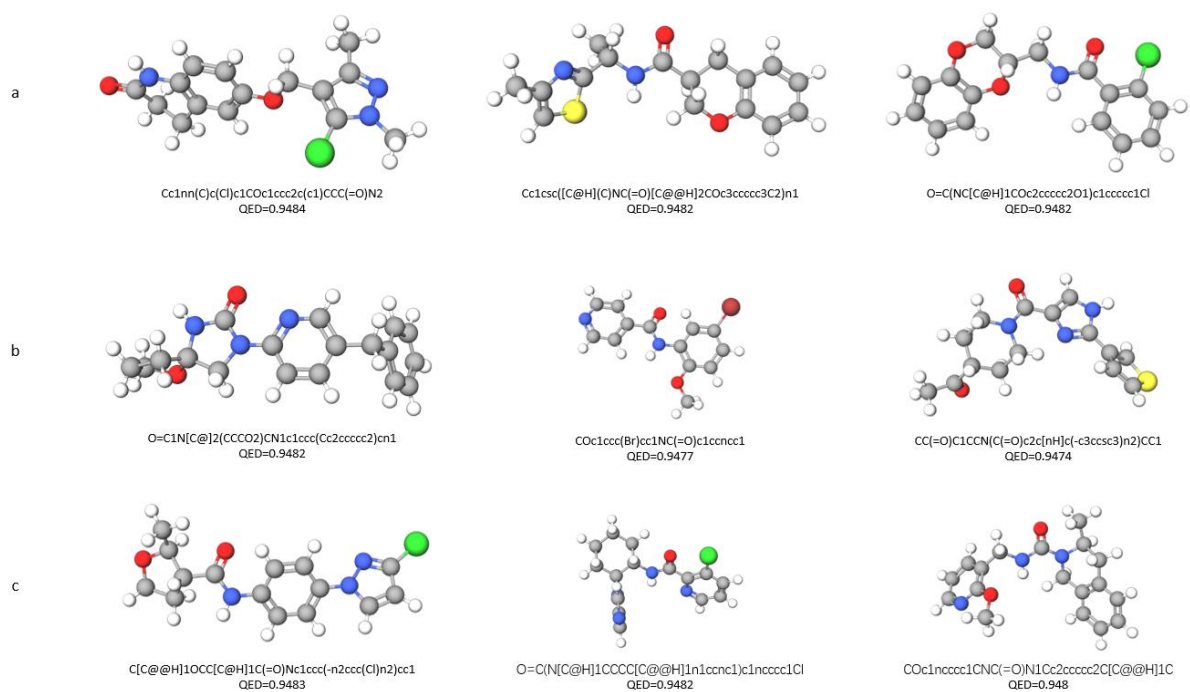

In the GG task, both 1 to N and N to 1 tasks employ atomic-level encoding, achieving good results in the QED targeted generation task. From the perspective of the QED values of generated molecules, the impact of pre-training methods on model performance is not significant. However, from the perspective of DG tasks, pre-training on N to 1 tasks still helps to enhance the validity of generating molecules.

To further visualize AdaMR's generative capabilities, we present the 3D structures of the top 3 generated molecules in **Figure 3**. Our model generates the top 3 molecules that not only closely match the target QED value but also exhibit high validity and stability. The structures display diversity, featuring various skeletal arrangements, which is beneficial for practical applications and further optimization in real-world scenarios.

The results of the property prediction tasks based on the two pre-training methods are shown in **Table 10**. We employ atomic-level molecular encoding in this experiment. From the perspective of classification ROC-AUC score and regression RMSE score, the N to 1 pre-training task is more effective than the 1 to N task on the property prediction task. The molecular property prediction task can be viewed as an induction of specific molecular properties in molecular space, while the N to 1 pre-training task can reduce the difficulty of the induction task by constructing a less complex molecular space than the 1 to N task on the same number of molecules. Therefore, for the property prediction task, the N to 1 pre-training task is more appropriate.

**Table 10**: Comparison of metrics for atomic-level encoding and substructure-level encoding in property prediction tasks based on different pre-training tasks.

| Task type | | AdaMR | |
|---|---|---|---|
| | | N to 1 task | 1 to N task |
| Classification (ROC-AUC) | BACE | 0.858 | 0.792 |
| | BBBP | 0.900 | 0.775 |
| | ClinTox | 0.961 | 0.856 |
| Regression (RMSE) | ESOL | 0.540 | 0.503 |
| | FreeSolv | 1.088 | 1.119 |
| | Lipophilicity | 0.776 | 0.790 |

# Conclusion

Bioinformatics has benefited greatly from the quick development of artificial intelligence technology. Pre-trained models are becoming the main focus of research and are being used more and more across various vertical fileds to support real-world applications. However, integration of multiple downstream tasks is generally required for practical implementations. Due to their fixed granularity of input encoding, existing pre-trained models are unable to fully capure information across various input levels. As a result, current pre-trained models face difficulties due to this discrepancy in information absorption across different input levels, which causes fluctuations in their performance across various downstream tasks that call for different information levels. Such disparities eventually have an impact on the efficacy of practical implementations.

In this study, we introduce a novel approach termed Adjustable Molecular Representation (AdaMR), comprising an adaptable molecular encoding scheme with adjustable granularity and a molecular canonicalization pre-training strategy for SMILES-based molecular pre-trained models. We employ a Transformer model to evaluate AdaMR's performance across several downstream tasks. Our experimental results demonstrate that our proposed model either matches or surpasses current SOTA methods across a range of downstream tasks. Furthermore, our study reveals the disparate demands of different downstream tasks concerning molecular encoding granularity. Substructure-level molecular representation proves advantageous for tasks such as property prediction, as it retains information about specific atom groups or arrangements crucial for determining chemical properties and functions. Meanwhile, atomic-level representation improves the validity, novelty, and uniqueness of generative tasks, particularly when combined with pre-training tasks utilizing generative molecular canonicalization. Based on these findings, we propose a novel strategy for selecting encoding methods tailored to downstream tasks in molecular pre-training. With this strategy, downstream tasks that require information from several scales, such as AI drug design, will be integrated into a unified normal form.

# Author Contributions

Y.D. and H.C. designed and performed the experiments. H.C., Z.Y., R.F., W.T., and P.X. prepared the figures and tables, and wrote the manuscript. Z.Y. helped conduct some computational analysis. J.Z. and Z.G. conceived, initiated, designed, and supervised this study.

# Acknowledgement

We geratfully acknowledge financial support from Nature Science Foundation of Jiangsu Province, Major Project (BK20222008).

# Code Availability

The model and codes of AdaMR are available at https://github.com/js-ish/MolTx/tree/paper.